\documentclass[12pt,preprint]{aastex}

\newcommand{\ex}[1]{\mbox{$\times 10^{#1}$}}

\newcommand{\kms}{{\rm km \:s^{-1}}}
\newcommand{\lsol}{\mbox{$L_{\odot}$}}

\newcommand{\msol}{\mbox{$M_{\odot}$}}

\shorttitle{The HII Region Gum~48d}
\shortauthors{Karr et al.}
 
\begin{document}
\bibliographystyle{astron}

\title{Gum~48d: an evolved HII region with ongoing star formation}

\author{J.L. Karr} \affil{Academia Sinica Institute of Astronomy and
Astrophysics, Taipei, Taiwan, ROC}

\author{P. Manoj } \affil{Department of Physics and Astronomy,
University of Rochester, NY}

\author{N. Ohashi} \affil{Academia Sinica Institute of Astronomy and
Astrophysics, Taipei, Taiwan, ROC}

\begin{abstract}

High mass star formation and the evolution of HII regions have a
substantial impact on the morphology and star formation history of
molecular clouds.  The HII region Gum~48d, located in the Centaurus
Arm at a distance of 3.5 kpc, is an old, well evolved HII region whose
ionizing stars have moved off the main sequence. As such, it
represents a phase in the evolution of HII regions that is less well
studied than the earlier, more energetic, main sequence phase. In this
paper we use multi-wavelength archive data from a variety of sources
to perform a detailed study of this interesting region.
Morphologically, Gum~48d displays a ring-like faint HII region
associated with diffuse emission from the associated PDR, and is
formed from part of a large, massive molecular cloud complex. There is
extensive ongoing star formation in the region, at scales ranging from
low to high mass, which is consistent with triggered star formation
scenarios.  We investigate the dynamical history and evolution of this
region, and conclude that the original HII region was once larger and
more energetic than the faint region currently seen. The proposed
history of this molecular cloud complex is one of multiple, linked
generations of star formation, over a period of 10 Myr. Gum~48d
differs significantly in morphology and star formation that the other
HII regions in the molecular cloud; these differences are likely the
result of the advanced age of the region, and its different
evolutionary status.

\end{abstract}


\section{Introduction \label{intro}}

High mass star formation has a profound impact on the interstellar
medium, through a combination of effects including ionizing radiation
and expanding HII regions, stellar winds and supernovae.  The effects
and appearance of high mass star formation vary dramatically throughout
the evolution of an OB star. When a high mass star begins burning
hydrogen, it produces significant amounts of UV radiation, which
immediately start to ionize the surrounding neutral and molecular
gas. At very early stages, the star is still heavily embedded in its
natal cloud, surrounded by a small (0.1 pc) ultra-compact HII region
(UCHII), and visible only at radio and infrared wavelengths. The
region then expands into a classic HII region.  At this phase of
evolution, the OB star(s) are surrounded by a hot bubble of ionized
gas, the Stromgren Sphere, initially a few parsecs in size, which due
to the pressure imbalance expands into the ambient medium and
compresses the surrounding molecular material.

The HII region, now several parsecs or more in size, is bright in both
radio (free-free and recombination line emission) and optical
(particularly H$\alpha$ emission). The heating of dust grains in the
vicinity of the region leads to bright emission from dust in the mid
to far-infrared, particularly in the photodissociation region (PDR)
between the ionized and molecular gas. If the star forming region
contains multiple OB stars, the HII region can evolve into a giant HII
region as individual HII regions combine to form an extended, often
irregularly shaped ionized region up to 100 pc in diameter. Ongoing
star formation is generally seen surrounding the HII region.

During the main sequence and later post-main sequence phase the
ionizing stars can experience significant mass loss in the form of
super-sonic stellar winds. These winds expand into the already ionized
HII region, creating a stellar wind bubble, and eventually deposit
their mechanical energy into the surrounding ISM \citep{arthur07b}. In
the final stages, supernova explosions from the OB stars can expand
into the combined HII region/stellar wind bubble, further disrupting
the remnants of the original molecular cloud \citep{garcia96}.

At the very end of the HII region evolutionary sequence only the
remnants of the violent disruptions of the molecular cloud remain;
shells and super shells of neutral material that have been swept up
through the combined effects of the HII region, stellar winds and
supernova remnants, eventually slowing down and fragmenting
\citep{tenorio88}. By this point, the presence of the initial high
mass ionizing stars is inferred from the shape of the ISM, as the
ionizing stars have long since progressed through the main sequence to
the later stages of evolution, to supernovae, to the point where even
the SNR are no longer readily observable.

The detailed effects of high mass star formation on the ISM and on
subsequent star formation activity are complex and contradictory. The
expanding HII region, stellar winds and supernovae will eventually
combine to disrupt the natal molecular cloud. At shorter timescales
there is a significant influx of energy into the cloud, increasing
turbulence as well as potentially truncating circumstellar disks.  On
the other hand, HII regions have long been thought to enhance star
formation in molecular clouds through triggered or sequential star
formation \citep{elmegreen98}. The expanding HII region can compress
existing overdensities in the molecular material, leading to
instability and collapse; the RDI or radiatively driven implosion
model \citep{lefloch94}. Alternatively, a shell of material can be
swept up around the expanding ionized region, eventually becoming
unstable, fragmenting and collapsing into new stars; the collect and
collapse model \citep{elmegreen77}.  Quantifying the effects of an HII
region of the star formation rate of a region (or even attempting to
'prove' a proposed example of triggered star formation), is
non-trivial, and in the case of an individual HII region often
inconclusive.

The classic HII region, ionized by one or more stars, is to date the
best studied of the various stages of evolution of the HII region, due
to its high brightness, relatively large size and the fraction of the
lifetime of an HII region spent in this phase. The UC/CHII phase is
relatively short (a few 10$^5$ years) compared to the classic/giant
HII region phase, which is on the order of the main sequence lifetime of
the ionizing stars: approximately 1-7 x 10$^6$ years depending on the
initial masses of the stars. 

The shorter lived post main sequence phase involves a change in the
effective spectral type of the star and a corresponding drop in the
total flux of ionizing photons compared to the main sequence,
resulting in an extended but fainter HII region in the radio and
optical. The later phases, including the SN phase, are transitory.
The final remnant shell phase is longer lived but observationally
elusive, due to the absence of strong ionizing radiation and the
complexity of the interstellar medium in neutral hydrogen, making the
positive identification of remnant shells difficult.

The HII region Gum~48d, located in the Centaurus Arm, presents an
interesting example of the post main sequence phase of a classic HII
region. The region is ionized by a B0Ip supergiant, HR~5171B, with a
companion G8~Ia supergiant, HR~5171A, a high luminosity, extremely
variable star (a.k.a. V766 Cen, HD 119796), of interest in its own
right \citep{humphreys71,vangenderen92}.  The region was initially
identified in the Gum and RCW catalogs of optical HII regions as
Gum~48d and RCW~80 respectively \citep{gum55,rodgers60}.  Some
subsequent studies identify it as RCW~80
\citep{georgelin88,humphreys71,saito01}, while others confuse it with
a radio bright supernova remnant to the south
\citep{whiteoak96,rakowski01,green04}.  The distance to the SNR region
has been variously estimated as 4 kpc, 5.4 kpc and 7.8 kpc
\citep{guseinov04}, and is most likely at a further distance than the
HII region.  For clarity we will keep the identification Gum~48d for
the HII region and RCW~80 for the SNR.

In this paper we use multi-wavelength archived survey data from
various sources to investigate Gum~48d. Multi-wavelength archive data
provide a powerful tool for the study of extended HII regions. The use
of multiple wavelengths allows us to probe the different components of
the ISM; from hot ionized gas to cold molecular gas to dust, as well
as tracing the population of embedded stars. Archived data, on the
other hand, permits the exploration of large regions of sky, and the
discovery and selection of interesting HII regions which would not
necessarily be readily selected for more specific, pointed
observations.  

In Section~\ref{data} we outline the sources and utility of the
archive data used. In Section~\ref{gum} we discuss the
multi-wavelength morphology, ionization and structure of the
region. In Section~\ref{stars} the star formation content at various
mass scales is investigated.  We then discuss the dynamic evolution
and history of the region, including its local history
(Section~\ref{history}) and speculate as to the possibility of
triggered star formation caused by Gum~48d
(Section~\ref{trigger}). Finally, the position of Gum~48d in the
larger scale structure and evolution of the region is discussed in
Section~\ref{lss}.

\section{Data \label{data}}

There are a number of archived data sets available for this
region. The region was observed as part of the Spitzer GLIMPSE Legacy
Survey of the inner galaxy, at 3.6, 4.8, 5.6 and 8 microns, all with a
resolution of about 1'.2 \citep{benjamin03}.  All the bands except 4.8
microns contain emission features from Polycyclic Aromatic
Hydrocarbons (PAHs), large, complex interstellar molecules. PAHs
absorb UV radiation and re-emit in various mid-infrared spectral
features. The 8 micron band, in particular, contains two of the
strongest PAH features, at 7.7 and 8.6 $\mu$m.  Emission from PAHs is
an excellent tracer of photodissociation regions (PDRs), the interface
between neutral and molecular hydrogen, as PAHs are destroyed by shock
fronts in the interior of ionized regions, and are effectively
shielded from UV radiation in dense molecular material.  The 4.8
micron band contains a rotational line of shocked molecular hydrogen,
while all four bands also contain thermal emission from small grains.

In addition to the GLIMPSE data we have used archive data from the
MIPS-GAL survey at 24 microns, a longer wavelength companion survey to
GLIMPSE. At this wavelength, the relevant emitter is small
dust grains, which may or may not correspond to the PAH morphology,
depending on the physical conditions in the region; PAH emission and
warm dust continuum generally coincide, but warm dust frequently exists
in the absence of PAH emission.

Point source catalogs are available for both the GLIMPSE data and
the 2 Micron All Sky Survey (2MASS), the later at J, H and K$_s$
\citep{skrutskie06}. We have performed point source extraction on the
24 micron data archive data (which are of good quality) and combined
the results with the preceding two catalogs to create 1-24 micron
spectral energy distributions (SEDs) for all point sources in the
region. The point source SEDs provide an excellent tracer for young
stellar objects (YSOs), due to the strong near and mid-infrared
emission from protostellar disks and envelopes, and the low extinction
at these wavelengths.

Optical images were obtained from the Southern H Alpha Sky Survey
Atlas (SHASSA) \citep{shassa}, with a resolution of 1'.  Radio
continuum (1420 MHz) and line (HI 21 cm) data sets at 1' were obtained
from the Southern Galactic Plane Survey (SGPS) \citep{sgps}. Both data
sets trace the morphology and brightness of the ionized gas, through
H$\alpha$ and free-free emission respectively. The optical data are
strongly affected by extinction, unlike the radio data, and so the
combination can provide a probe of three dimensional HII region
morphology.

\section{Gum~48d \label{gum}}

Figure~\ref{3col} shows the large scale structure of the region
surrounding Gum~48d on a two degree scale; 1420 MHz emission, tracing
ionized gas in HII regions and SNRs, is shown in blue, GLIMPSE 8
micron data, tracing PAH emission PDRs, is shown in green, and 24
micron MIPSGAL data, tracing warm dust, is shown in red. The contours
show the location of molecular gas ($^{13}$CO(J=1-0)) in the velocity
range of the Centaurus Arm, from the survey of
\citet{saito01}. Individual HII regions in the area are labeled, while
the region containing Gum~48d is outlined with a box.

The emission from ionized gas is shown in Figure~\ref{ion}, in optical
(left panel) and radio (right panel).  The optical emission in
H$\alpha$ shows a faint, semi-circular nebulosity surrounding the
central star.  The central star seen in the image is HR~5171A, the
high luminosity G8~Ia supergiant which dominates the optical and mid
infrared images. It is a binary star, whose companion, HR~5171B, is a
much fainter B0~Ip supergiant and source of ionization for the
region. The two stars have a kinematic distance of 3.9 kpc and a
photometric distance of 3.5 kpc \citep{humphreys71}.  If we place the
two stars on an HR diagram evolutionary track, using the models of
\citet{schaller92}, HR 5171A has an age of 3.5 Myr and a main sequence
mass of 60 $\msol$, while HR 5171B has an age of 4 Myr and a main
sequence mass of 40 $\msol$, corresponding to main sequence spectral
types of O7 and O5.5 respectively \citep{schaerer97}. This is
consistent with the conclusions of \citet{vangenderen92}. This implies
that Gum~48d was once a very energetic HII region, ionized by multiple
massive O stars.  The above studies of this region identify only two
stars in this system. An inspection of the Tycho catalog shows two
stars at the location of HR~5171B with similar optical magnitudes,
which might indicate the possibility of an additional star. No
spectral information is available to confirm this, but the ionization
balance calculations shown in Section~\ref{history} rule out the
need for  an additional ionizing source. We will discuss the
evolutionary history of Gum~48d in more detail in
Section~\ref{history}.

The optical morphology is mirrored in the radio data.  The radio image
is dominated by the SNR RCW~80 to the south-west of Gum~48d.  The
ionized gas of the HII region has measured velocities of -48 km~s$^{-1}$
\citep{georgelin88}, corresponding to a near kinematic distance of 4
kpc, using the galactic rotation model of \citet{brand93}. This value
is in good agreement with the distances to the central stars.  There
are five other 1420 MHz features in the region of
Figure~\ref{ion}. Gal 309.54-0.737 is an HII region with a velocity of
-43 km~s$^{-1}$ \citep{caswell81}.  The other four are compact 1420 MHz
sources without known velocities. G309.14-0.18 is a small but
extended HII region with a spectral index consistent with an HII
region \citep{caswell92}, while G309.17-0.02, G309.27-0.03 and
G309.24-0.18 have no spectral index measurements.  All four sources
are positionally coincident with extended emission at 24 and/or 8
microns, as described in the following section.

\subsection{Dust and Gas \label{molec}}

In the mid-infrared Gum~48d has a striking morphology, as seen in
Figure~\ref{pah}. The main feature is an extended, bow shaped region
of reasonable thickness compared to its full extent, and centered on a
bright infrared point source (HR~5171A). The morphology within the
region is complex and detailed on smaller scales, with significant
substructure. There is a small, bright region to the very south-east
of the image, corresponding to the HII region Gal 309.54-0.737, as well
as a number of bright, compact mid-IR features scattered along the bow
shaped region.  There are also small, arc-shaped structures, most
noticeably one to the east of the center of the image.  Finally, there
are several infrared dark clouds, seen in extinction (areas of no
emission) in the 8 micron map, silhouetted against brighter emission,
and a large number of infrared point sources.


This morphology is mirrored at 24 microns, as seen in
Figure~\ref{warm}. In this image, tracing the location of warm dust,
the substructure and compact features are even more prominent than at
eight microns. The locations of the compact features (discussed more
fully in Sections~\ref{compact} and \ref{uchii}) are marked in
Figure~\ref{warm}.  Several of the infrared dark clouds persist at 24
microns, while others are now seen in emission. There are two
saturated regions in the 24 micron image, one corresponding to
HR~5171B (the central region) and one to the AGB star GLMP~363, to the
south west. Post main sequence stars are often extremely bright in the
mid-infrared, due to the presence of surrounding dust.

\citet{saito01} conducted $^{13}$CO(J=1-0) observations of this region,
identifying an extensive molecular cloud at velocities corresponding
to the Centaurus Arm of the galaxy and a kinematic distance of 3.2 -
5.5 kpc, with a total mass of 2.3$\ex{6}$ $\msol$; an extremely
massive molecular complex. The whole molecular cloud observed by
\citet{saito01} is shown in Figure~\ref{3col}. The molecular cloud
forms an extensive ``ridge'' perpendicular to the Galactic plane and
with a physical extent of 80 pc. This molecular ridge curves slightly,
and Gum~48d is located at its inner edge.

\citet{saito01} also carried out C$^{18}$O(J=1-0) observations in the
same region to investigate more detailed structures of the molecular
cloud, identifying several dense molecular cores.  The locations of
the C$^{18}$O molecular cores are marked in Figure~\ref{pah}; diamonds
mark the location of cores with velocities corresponding to Gum~48d,
crosses mark cores at more negative velocities.  The C$^{18}$O is
arrayed around the edge of the 8 micron PAH emission, outside of the
emission from the ionized gas, with velocities ranging from -40.6 to
-50.2 km~s$^{-1}$. This demonstrates that the bulk of the 8 micron emission
occurs at the interface between the ionized region and the molecular
cloud. In addition, it can readily be seen that the locations of the
infrared dark clouds are coincident with the positions of the C$^{18}$O
dense cores. These infrared dark clouds are very dense, often
filamentary clouds of molecular material that are optically thick in
the mid-infrared \citep{carey98}, and are sometimes associated with
highly embedded, very young protostars. We note that the smaller
arclike 8 micron feature located at 309.4 -0.37 is associated with
molecular gas at -60 km~s$^{-1}$. This region is either not physically
associated with Gum~48d, or it is a located at the front of the HII
region cavity, and is moving towards us with respect to the central
velocity of the ionized gas.

The HII region Gal 309.54-0.737 is associated with a molecular cloud
at -52.6 km~s$^{-1}$.  Its morphology at 24 and 8 microns, when
compared with the 1420 MHz emission and location of the molecular
cloud, is indicative of a compact blister HII region, sharply bounded
to the south east and venting to the north-west.  The region shows a
strong alignment between the location of the emission of ionized gas
and mid-infrared emission from dust. The nebula is only very faintly
seen in the optical, at the unbounded north-west edge, corresponding
to the fainter radio emission of the venting gas.

The other four compact HII features are all associated with features
at 24 and 8 microns.  G309.14-0.18 is associated with an infrared dark
cloud at 8 microns, but with clumpy emission at 24 microns, brighter
to the western edge (denoted S10 in Figure~\ref{warm}). G309.17-0.02 is
associated with compact emission at both 24 and 8 microns (denoted S11
in Figure~\ref{warm}), while G309.27-0.03 is associated with compact
emission at only 24 microns (denoted S12 in
Figure~\ref{warm}). G309.24-0.18 is located near an arc like structure
at 8 microns and faint interior emission at 24 microns (denoted S4 in
Figure~\ref{warm}). The molecular emission components are not well
aligned, however, and most likely are the result of a coincidental
projection onto G309.24-0.18.

The velocities of the molecular material are consistent with the
published values for the ionized material, and have  kinematic
distances consistent with the photometric distance of the ionizing
star.  That, combined with the morphological correspondence between
the dust, molecular and ionized emission, indicates that all observed
components of the region are located at a distance of ~3.5 kpc and are
physically associated with each other.

\section{Star Formation \label{stars}}

Gum~48d shows extensive, ongoing star formation over a range of
masses. Low and intermediate mass star formation is well traced in the
near to mid-infrared, with several methods of identifying young stars
based on infrared photometry available. The first is by color; at
IRAC wavelengths young stars show redder colors, with more embedded
sources being progressively more red \citep{allen04}. At these
wavelengths, extinction is minimal and shows little variations with
wavelength \citep{indebetouw05}. A complementary technique is the
measurement of the mid-infrared spectral slope, $\alpha$, using the
definitions of \citet{adams88}. Sources with rising mid-IR slopes are
identified as Class I sources, young stars still heavily embedded in
their molecular envelope. Those with negative slopes that are
shallower than photospheric emission are Class II sources (Classical
T-Tauri and Herbig Ae/Be stars), which are less embedded but still
have a significant circumstellar disk, and possibly remnant
envelope. An additional category, flat-spectrum sources, straddles the
boundary between the two. The even younger Class 0 sources, are not
well identified by this method, due to the extreme extinction at
shorter wavelengths.

There are a total of 48411 2MASS sources, 39176 GLIMPSE catalog
sources and 517 extracted MIPS 24 micron sources in the region shown
in Figure~\ref{pah}. The three catalogs were combined to produce a
single band merged catalog; this final catalog contains 76696
sources.  Of these, 12670 have sufficient fluxes for an IRAC
color-color diagram (i.e. fluxes are measured in all four IRAC bands).
The IRAC color-color diagram is shown in Figure~\ref{cc1}. The
sources at 0,0 are photospheric SEDs, with more embedded (redder)
sources occurring to the upper right.

We also calculate the mid-infrared spectral slopes for all sources
with photometry in at least four bands between 2MASS K$_s$ and IRAC 8
microns (21284 sources), and identify Class I, Flat Spectrum and Class
II sources, for a total of 1827 sources.  The classification based on
the mid-infrared SED slope is, in general, very well correlated with
the classification via IRAC colors, however the method is more robust
in the presence of a missing IRAC flux.  Figure~\ref{sform} shows the
distribution of young stars identified above over the region.  In the
upper panel Class I sources are marked with crosses and flat spectrum
sources with diamonds, shown over the IRAC 8 micron image and tracing
the most embedded phase of star formation. In the lower panel the
location of Class II sources are shown with triangles.
 
There are several sources of contamination in method for identifying
potential YSOs; foreground and background galactic and extragalactic
sources. The high levels of extended emission in the mid-infrared
minimizes the contamination by faint background sources, either
galactic or extragalactic, as the detection limits at 24 and 8 microns
are generally limited by background levels rather than sensitivity to
the point source flux.  Unresolved star forming galaxies have IRAC
colors indistinguishable from young stars; however, the location of
this region in the Galactic plane, combined with the low flux of many
of these sources, minimizes this contamination.  For comparison, we
took a nearby region in the sky, of similar galactic latitude, 
without an HII region in either the optical or radio, and without
bright features in the mid-infrared, and determined the number of
sources with YSO colors. This gives us an estimate of the line of
sight contamination of sources. The region from l=309.56 to 310 and
l=-0.273 to -0.705 was the largest potential region; if we scale the
YSO counts to match the angular size of the Figure~\ref{sform} the
expected contributions from line of sight contaminants are 2\%, 14\%,
and 29\% for Class I, Flat Spectrum and Class II sources respectively.

At the distance of Gum~48d the 2MASS fluxes are of limited usefulness,
due to the typical limiting magnitude of the survey. Only the
brightest and/or reddest of the young stars at 3.5 kpc will be
detected, due to both the intrinsic luminosity and the high extinction
in the Galactic plane. The IRAC and MIPS data are considerably more
sensitive to objects with the SEDs of YSOs and are much less affected
by extinction.  K$_s$ band fluxes are the most commonly detected, and
are useful for the calculation of the spectral slope in cases where
one of the IRAC fluxes is missing.

In this region the Class II sources are primarily correlated with the
large molecular ridge, indicating ongoing star formation throughout
the cloud, even though Class II sources have a higher level of
contamination through background or foreground sources. The more
heavily embedded Class I and flat spectrum sources, however, are more
strongly clustered. There are several clusters or small aggregates of
very embedded young stars, many of which are associated with dense
cores identified in C$^{18}$O(J=1-0). The locations of the clusters,
determined by eye, are marked in the first panel of
Figure~\ref{sform}.  A cluster at 309.42-0.64 is associated with a
clump with a central velocity -42.4 $\kms$, one at 309.53-0.75 with
central velocity of -52.6 $\kms$, one at 309.22-0.47 with a clump with
a central velocity of -40.6 $\kms$, and at 309.14-0.19 with material
at -46.4 km~s$^{-1}$. A cluster at 309.38-0.14 is associated both with a
molecular clump with a velocity of -50.5 $\kms$ and with a maser (a
star formation tracer) at -50 $\kms$ \citep{caswell95}.  All of these
velocities are consistent with the velocities of the HII region and
its constituent components. A final cluster at 309.54-0.12 is not
associated with a high density region, although it is superposed on
extended $^{13}$CO emission.  From this distribution it is apparent
that star formation is continuing throughout the molecular ridge, and
that the most embedded phases of star formation are preferentially
occurring in the densest molecular gas; many of these clusters are
also associated with IR dark clouds.

Age estimates of young stars are difficult to derive, as stars of
different masses evolve at different rates, and even stars of similar
masses can show a range of different spectral slopes for a given age,
due in part to the effects of orientation \citep{robitaille07}. The
general consensus, however, is that rising and flat spectrum sources
are, on average, a younger and more embedded phase of star formation
\citep{allen04}. Thus, the younger population of young stars is more
highly clustered than the older population.

\subsection{Intermediate Mass Star Formation \label{compact}}

In addition to the IRAC point sources tracing primarily low and
intermediate mass star formation, there is evidence for ongoing star
formation at higher masses.  There are a number of infrared features
throughout this region, which share the common properties of being
bright, compact (less than a few arcseconds) and clearly non-point
sources while at the same time not being diffuse.  Some of the sources
are associated with emission from ionized gas, while others are not.
We will divide the sample into two groups; intermediate and high mass
star formation, based on the absence or presence  of emission from
ionized gas. Intermediate mass star formation is discussed below,
while high mass star formation is discussed in Section~\ref{uchii}.

The first sample of infrared sources shows compact but distinctly
non-point like morphology at 24 microns, and compact but structured
morphology at 8 microns. Figure~\ref{compact1} shows a gallery of these
sources, while Table~\ref{compactparms} lists the positions, angular
sizes, physical sizes (assuming a distance of 3.5 kpc) and 8 and 24
micron fluxes.  The 24 micron flux is in general smoothly distributed,
even considering the larger 24 micron PSF, but occupies the same
physical extent as the more structured 8 micron emission. To confirm
this, we convolved the 8 micron data with the 24 micron PSF for
comparison.  These sources do not show an obvious 24 or 8 micron
embedded point source, and show no evidence of compact or point
like 1420 MHz emission at the sensitivity of the SGPS survey. The
diameters of the sources range from 0.5 - 2 pc.

A possible explanation for these sources is that they are enshrouded
intermediate mass stars \citep{fuente98, fuente02, roger92,
diazmiller98}.  The lack of radio continuum emission associated with
these sources argues against the presence of an UCHII region and
therefore rules out the presence of stars earlier than type B1, as a
UCHII region around more massive stars at this distance would be
visible at the sensitivity of the radio data.  On the other hand,
bright, extended 8 micron emission is often indicative of a
photodissociation region, which can be created by sub-Lyman
ultraviolet radiation. The 24 micron emission traces warm, small
grained dust, which can also be heated by sub-Lyman photons.
Intermediate mass stars of type later that B1 could satisfy these
irradiation conditions, while stars lower in mass than B8 would not
provide enough ultraviolet radiation to illuminated a strong PDR. The
lack of a clearly defined point source within the region could be due
to sensitivity, extinction, or the high level of mid-IR emission
making point sources difficult to detangle from the structured,
extended emission, a particular problem at 24 microns.

Although there are no distinguishable point sources within the compact
features, there is a spatial correlation between the compact features
and the embedded young stars, which strengthens the claim that they
are associated with ongoing star formation.  Sources S7 and S8 are
associated with a small aggregate of deeply embedded sources, as well
as a maser source, while S3 is at the edge of another small
aggregate. Sources S5 and S6 are associated with the star formation at
the inner edge of the HII region, and S1 and S2 are associated with
another concentration of embedded star formation. Source S9 is located
near the heavily embedded star formation of Gal~309.54-0.737, as well
as a small cluster of highly embedded sources. These positional
coincidences indicate that these sources are physically related to
star formation in this region; they are unlikely to be fore or
background projections.

\citet{kerton02} identified 165 potential candidates similar to those
described above based on CO data in the Canadian Galactic Plane
Survey and MSX 8 micron survey data, with a limiting distance of 2
kpc, indicating that these might not be particularly rare objects. The
objects shown here are of comparable size (1-2 pc in diameter) to the
objects in the MSX sample, albeit of smaller angular size.  We
computed IRAS luminosities and colors for those sources with IRAS
counterparts (S1, S2, S3, S7, S8, S10 and S11), as tabulated in
Table~\ref{compactparms}. The luminosities, sizes and colors are all
within the range found by \citet{kerton02}. We attempted to duplicate
the HI velocity analysis performed therein, but did not have
sufficient spatial resolution in the 21 cm data to reach a conclusion.  The
SGPS data are of similar resolution to the previous study, but the
sources are more distant by a factor of at least two than those
sources.

It should be noted that at a distance of 3.5 kpc a B0 star would be
expected to have a visual magnitude of at least 14.7, and a B5 star
18.0, using the standard value of 1.6 magnitude of extinction per kpc
in the Galactic plane. The actual magnitude of an enshrouded B star
could be much greater due to the high density cocoon of material.
Therefore, we are unlikely to detect an enshrouded B star at this
distance in the available survey data.  High sensitivity NIR
observations could provide a probe of the embedded B star and
potential lower mass cluster.

Sources S7 and S8 show a particularly interesting morphology; a
vaguely bi-polar structure (two lobed) with several deeply embedded
sources (24 micron bright) in between.  The morphology is reminiscent
of a bi-polar outflow with a wide opening angle.  The SEDs and 8 and
24 micron morphologies are similar of those of the other compact
sources which have been identified as possible enshrouded B
stars. This might be an outflow near an enshrouded star, or could
possible be a pair of enshrouded B stars in a cluster of young
stars. The lack of bright IRAC Band 2 data (tracing molecular shocks)
argues against the first option, as do the SEDs. Therefore, these are
most likely to be enshrouded B stars.

It is worth considering the likelihood of observing an intermediate
mass B star in this phase, as these are not widely observed
objects. Estimates of the ages of this class of objects vary, however,
consensus is that they are less than 10$^6$ years old \citep{fuente98,
fuente02, roger92,diazmiller98}. The general lack of associated
masers indicates a relatively evolved status \citep{kerton02}. This is
a longer timescale than the typical estimated ages of UCHII regions
(up to a few 10$^5$ years), a much better studied class of object. A
young, enshrouded B star would not be identified through radio
surveys, the typical method of discovery for UCHII regions, and mid-IR
data suitable for a general search have only recently become generally
available.  Furthermore, an enshrouded B star would only be detected
with this morphology under the right conditions. In a lower density
environment the B star could more easily break out of its cocoon,
while in the close presence of ionizing stars the observational
signature of the B star would be overwhelmed. Therefore, the
relative lack of previous studies of these objects is likely the
result of both the difficulty of detecting them in previously
available data and the combination of factors required to make them detectable
in the infrared.

\subsection{High Mass Star Formation \label{uchii}}

There is further evidence for ongoing high mass star formation in this
region, as traced by emission from ionized gas. The infrared
morphologies of the high mass star formation candidates are shown in
Figure~\ref{compact2} (S9, S10, S11 and S12; see also
Table~\ref{compactparms}), while the HII regions themselves are
labeled in Figure~\ref{ion}, with additional information in
Table~\ref{compactparms}.  In general, the high mass star formation
candidates show more extended emission in the infrared than the
intermediate mass candidates. The first source, S10, is associated
with the HII region, G309.14-0.18 which coincides with an IR-dark
cloud at 8 microns, a bright arc along the west side and bright,
clumpy emission (including several bright point sources) at 24
microns.  The IR-dark cloud is associated with one of the C$^{18}$O
clumps observed by \citet{saito01}, with a velocity of -46.4
km~s$^{-1}$ and with a cluster of embedded young stars, as discussed
above.  The 1420 MHz radio emission has a spectral index consistent
with an HII region and is resolved in the 1' resolution SGPS data. The
ionizing star is not known and is undetectable at optical or mid-IR
wavelengths due to the high foreground extinction.

To the south of Gum~48d there is another high mass star formation
region, Gal~309.54-0.737, a blister HII region. This HII region is
clearly identified as an HII region by its spectral index, although
the ionizing star is not seen due to high extinction.  The velocity of
the C$^{18}$O emission associated with this region (-52.6 km~s$^{-1}$)
corresponds to the Centaurus Arm, although \citet{saito01} place it at
the far kinematic distance, based on the lack of associated optical
emission.  However, the region is actually visible in the H$\alpha$ data
in the northwest (corresponding with the direction of the champagne
flow), although it is extremely faint.  Therefore it is likely that
the faintness of the region in the optical is due to strong local
extinction, rather than the region being at a significantly larger
distance than the other HII regions in the area.

Though the ionizing stars associated with these two compact HII
regions are unknown, we can use the 1420 MHz flux to estimate the
spectral type of the star required to produce the ionizing flux
necessary to maintain the observed free-free emission, using the radio
free-free spectral luminosity

\begin{equation}
\frac{{N}_{\rm{Ly}}}{\rm{s}^{-1}} \approx 6.3 \times 10^{52}(
\frac{{T_e}}{10^4 \rm{K}})^{-0.45} (\frac{\nu}{\rm{GHz}})^{0.1}
(\frac{{L}_{\nu}}{10^{20}\rm{W\ Hz^{-1}}})^{-1}
\label{lyman}
\end{equation}

where $N_{Ly}$ is the flux of ionizing photons, $T_e$ is the electron
temperature and $L_{\nu}$ is the observed luminosity at frequency
$\nu$.  This calculation has several uncertainties. It assumes
ionization equilibrium. The calculated value will be a lower limit, as
a matter bounded HII region will leak ionizing flux into the
surrounding lower density medium. In addition, the ionizing flux could
be the result of multiple, lower mass stars rather than a single,
higher mass star.  Gal 309.54-0.737 has a background subtracted flux
of 1.6 Jy at 1420 MHz, a distance of 3.5 kpc and an electron
temperature of 5900 K \citep{caswell87}.  These values give a Lyman
flux of 1.5$\ex{47}$, corresponding to a main sequence spectral type
of B0.5 \citep{vacca96}. The calculated Lyman flux of G309.14-0.08 is
below the model values for OB stars, indicative of its young age.
This implies that these HII regions are ionized by single stars,
unless significant ionizing flux is escaping from these regions. This is
unlikely due to the extremely dense surrounding material.

There are two further HII regions with mid-infrared counterparts.
The compact HII regions G309.17-0.02 is associated with a compact
feature at 8 and 24 microns (S11) and is morphologically similar to
the B star candidates, but its spectral index and velocity are not
known. The region at G309.27-0.03 is associated with compact
emission at 24 microns only (S12), although there is a faint arc like
8 micron feature at its edge.

To summarize the star forming properties of this region:

\begin{itemize}

\item There is extensive, ongoing, low mass star formation associated
with the molecular material surrounding Gum~48d. 

\item The more embedded and younger phases of star formation are more
strongly clustered than the more evolved Class II sources, and are
strongly correlated with the densest molecular material.

\item There is evidence for intermediate mass star formation, in the
form of potential enshrouded B1-B8 stars. These sources are also
correlated with the location of molecular material and low mass star
formation.

\item There are five UC-small HII regions in the area, indicating high
mass star formation. Four of them are  positively identified with
either a cluster of embedded young stars, or molecular material at
Centaurus arm velocities, indicating a physical association with
Gum~48d.

\end{itemize}

\section{Discussion \label{discuss}}

In the following discussion, we attempt to describe the dynamical and
star formation history of Gum~48d, and its place in the large scale
structure and history of the Centaurus arm region. In
Section~\ref{history} we describe the dynamical history of Gum~48d,
using known physical parameters, and compare it with values
predicted by models for the expansion of an HII region. In
Section~\ref{trigger} the proposed history and the current observed
star formation activity are compared to evaluate the region in the
context of triggered star formation. Finally, in Section~\ref{lss}
Gum~48d is compared with other HII regions in the area, and its place
in the large scale structure and star formation history of the
Centaurus arm is analyzed. 

\subsection{History of Gum~48d \label{history}}

In this subsection, we attempt to describe the history of Gum~48d,
illustrated in panels (a) to (d) in Fig. 9. We then discuss
Gum~48d's atypical appearance when compared to other nearby HII
regions in the context of this history.

As briefly discussed in Section~\ref{gum}, Gum 48d is currently
ionized by HR 5171B, a B0 Ip post-main sequence star. This is fully
consistent with a spectral type estimation of the ionizing star using
Equation~\ref{lyman} and the 1420 MHz radio flux of Gum 48d. We should
note, however, that the duration of the post-main sequence phase is
short compared to the main sequence lifetime of the ionizing stars,
suggesting that the dynamical evolution of Gum 48d and its expansion
into the surrounding ambient material will have been mostly determined
when HR 5171 B and its companion, HR~5171A, were in the main sequence
phase, with spectral type of O5.5 and O7, respectively (see Section
~\ref{gum}).

How much did Gum 48d extend into the surrounding ambient material when
the two ionizing stars were in the main sequence phase? The expansion
of an HII region into a uniform density ambient medium can be simply
described by the model of \citet{dyson97}, in which

\begin{equation}
\frac{R}{\rm{pc}} \approx 4.5\ \ 
(\frac{N_{\rm{Ly}}}{10^{49}  \rm{s}^{-1}})^{1/7}
(\frac{n}{10^3 \rm{cm}^{-3}})^{-2/7}
(\frac{t}{\rm{Myr}})^{4/7}
\label{dyson} \end{equation}

\noindent $R$ is the radius of the HII region, $N$ is the flux of
ionizing photons, $n$ is the ambient density of the surrounding
material, and $t$ is the time scale of the expansion, which should be
equivalent to the age of the ionizing star.  Based on the main
sequence masses and spectral types of HR 5171A and B, estimated using
an HR diagram in Section~\ref{gum}, the time scale of the expansion
and the ionizing flux during the main sequence phase are 3.8 Myr and
1.1$\times 10^{50}$ photons s$^{-1}$, respectively. If we assume that
the radius of Gum 48d during its main sequence phase is the same as
its current size, 8.4 pc measured from the radio free-free emission at
1420 MHz, the estimated ambient density of the original molecular
cloud is 5.7$\times 10^3$ cm$^{-3}$, an unreasonably high value for a
molecular cloud. For comparison, the average density of the current
molecular ridge is ~500 cm$^{-3}$, estimated from its averaged column
density, 1.9 $\times 10^{22}$ cm$^{-2}$ \citep{saito01}, assuming that
the molecular ridge is of a similar physical scale along the line of
sight as it is in width.

There are several possible explanations for the discrepancy between
the calculated value for the density and physically plausible values.
One scenario is blow-out; if the HII region breaks out of the ambient
material into a lower density region, a champagne flow will
occur. This will slow the expansion of the HII region into the denser
medium \citep{tenorio79}. Another possibility is that the HII region
created during its main sequence phase was larger than the currently
ionized region. This makes more sense physically because the
difference between the ionizing fluxes during the main sequence and
post main sequence phases is substantial, and we would expect a more
energetic HII region in the past. For example, the W5-West HII region,
with an ionizing flux half that of Gum~48d during its main sequence
phase, has a current radius of 35 pc \citep{karr03a}. Similarly, the
wind blown bubble models of \citet{arthur07b} for a 40 solar mass star
predict a final radius at the end of the stellar wind phase of on
order of 30 pc.

How do we recognize the maximum size of Gum 48d during the main
sequence phase? One potential way is to use the PAH emission at 8
microns arising from the PDR region, which occurs at the interface
between the molecular and neutral gas, and can be excited by sub-Lyman
UV photons. Consequently, it is a better tracer of the edge of the
molecular shell than the emission from ionized gas.  The material
surrounding the HII region forms a roughly semi-circular shell,
illuminated by the PAH emission. The molecular ridge is significantly
larger in scale than the PAH shell around Gum~48d, as is indicated in
panel (d) of Figure~\ref{sketch}. 

The recombination timescale for ionized hydrogen is approximately
10$^{5}$/$n$ years, where $n$ is the number density of the gas. The
timescale for the formation of molecular hydrogen from atomic hydrogen
is approximately 10$^{9}$/$n$ \citep{tielens85}. For a typical HII
region density of 10 cm$^{-3}$ this corresponds to 10,000 and 10$^8$
years respectively. Therefore, in the time since HR~5171B has left the
main sequence and the ionizing flux has dropped, there has been
sufficient time for the ionized region to recombine into atomic
hydrogen, but not to re-form molecular hydrogen, leaving an extended
neutral envelope around the current HII region. 

If we take this PAH shell as the boundary of the original HII region,
the radius of the HII region is roughly estimated to be 15 pc when
measured across the carved out semi-shell. The ambient density
calculated via the model above is now 840 cm$^{-3}$, approaching
physically plausible values. It should be noted that the HII region is
not spherically symmetric, and the radius estimated changes somewhat
with the assumed geometry of the HII region. The assumed shape and
center of the PAH shell were chosen as an intermediate estimate of the
size of the partial shell.

The current structure and appearance of Gum~48d therefore seem to be
the result of its evolutionary status. The shape of the surrounding
molecular/neutral material is dominated by the large scale structure
of the molecular ridge and the actions of the energetic main sequence
phase, while the current ionization state is the product of the post
main sequence star, HR~5171B. Gum~48d differs significantly in
appearance when compared with nearby, more typical HII regions, where
the ionized region borders directly on a bright PDR (as discussed
further in Section~\ref{lss}).  Unlike these regions, the PDR and the
edge of the ionized regions are not congruent in Gum~48d, and the PDR
is fainter and more diffuse.

\subsection{Triggered Star Formation \label{trigger}}

When studying HII regions and ongoing star formation, the question of
triggered star formation inevitably arises. Is the ongoing star
formation a consequence of the effects of the expanding HII region on
the surrounding ISM?  Proving this conclusively is problematic at
best. Typical methods involve 'not disproving' the hypothesis, often
by assembling a consistent time line for the different phases of star
formation. 

The best case for triggered star formation in this region, given its
age and morphology, is the collect and collapse model, where the
expanding HII region compresses the surrounding material and sweeps up
a shell of dense gas. At some point, the dense shell becomes unstable,
fragments, and collapses to form stars \citep{elmegreen77}.  In order
for a region to be consistent with a triggered star formation
scenario, the time from the formation of the ionizing star and turning
on of the associated HII region to the gravitational instability and
fragmentation of the swept up shell, plus the age of the putatively
triggered star formation population must be comparable to the total
age of the region.

As discussed earlier, the age of Gum~48d is $\simeq$4 Myr, based on
the ages of the ionizing stars. The age of the YSO population is not
easily calculated, and it is not possible to perform such a
calculation from the mid-infrared SEDs alone.  However, a typical age
for embedded low mass YSOs is on the order of or less than 1~Myr.  It
is difficult to assign a precise age for the embedded B star phase of
intermediate mass star formation, but their lifespan is estimated at
$<$ 1 Myr years \citep{dewdney91,roger92}. The age of UCHII regions
are also not well understood, largely due to uncertainties in modeling
the early stage of expansion, but they are a similarly young phase of
star formation (a few times 10$^{5}$ years \citep{garcia96}). To
summarize, the ages of the various populations of young stars at
different masses are on order of 1~Myr or less.

This leaves the epoch of the fragmentation of the shell to be
determined. This can be estimated via a simple analytical model.
Following the equations of \citet{whitworth94a}, which progress from the
expansion model of \citet{dyson97} used earlier, a shell created around an
expanding HII region becomes unstable to fragmentation and collapse at
a time

\begin{equation}
\frac{t_{frag}}{\rm{Myr}} \approx 1.56\ 
(\frac{a_s}{0.2 \rm{km\ s}^{-1}})^{7/11}
(\frac{N_{\rm{Ly}}}{10^{49} \rm{s}^{-1}})^{-1/11}
(\frac{n}{10^3 \rm{cm}^{-3}})^{-5/11}
\label{tfrag}
\end{equation}

\noindent $a_{s}$ is the sound speed (0.2 km~s$^{-1}$ an extreme lower
limit), $n$ is the number density of the ambient medium (10$^3$
cm$^{-3}$ an extreme upper limit) and $N_{Ly}$ is the flux of ionizing
photons. At this point the HII region and its corresponding shell have
reached a radius

\begin{equation}
\frac{R_{frag}}{\rm{pc}}\approx 5.8\ 
(\frac{a_s}{0.2  \rm{km\ s}^{-1}})^{4/11}
(\frac{N_{\rm{Ly}}}{10^{49} \rm{s}^{-1}})^{1/11}
(\frac{n}{10^3 \rm{cm}^{-3}})^{6/11}
\label{rfrag}
\end{equation}

\noindent with the variables as described above. Using the estimated
size of the original HII region calculated from the molecular/PAH
shell (15 pc) we can then work backwards; solving for the initial
density from the current size via Equation~\ref{rfrag}, and then
calculating the fragmentation timescale from Equation~\ref{tfrag}.
For a sound speed $a_{s}=0.5$ km s$^{-1}$ this gives an estimate of
the fragmentation timescale of 3.1 Myr and an initial ambient density
of 490 cm$^{-3}$, comparable to the current density of the extended
molecular ridge. The fragmentation timescale could be decreased if the
ionizing stars were to depart from the main sequence, which
corresponds to a decrease in ionizing radiation and subsequent cooling
of the swept up shell.

This time scale for fragmentation, plus the ages of the various
populations of young stars of different masses (about 1~Myr), is
comparable in age to the original HII region. Therefore, the star
formation can be explained as the result of triggering by the original
HII region.  The locations of the enshrouded B stars are significant in
this scenario, as they are located around the larger PAH shell
(corresponding to the original HII region) rather than the smaller,
currently ionized region. This is also consistent with formation via
triggering and the collect and collapse model.

More detailed modeling of Gum~48d would be problematic due to the
non-static large scale structure. The HII region has formed and
evolved in an expanding, increasingly more massive giant ridge of
molecular material.  This type of dynamically changing environment in
the surrounding ISM and its interaction with an evolving HII region is
beyond the scope of current models.

\subsection{Large Scale Structure \label{lss}}

\begin{figure}
\plotone{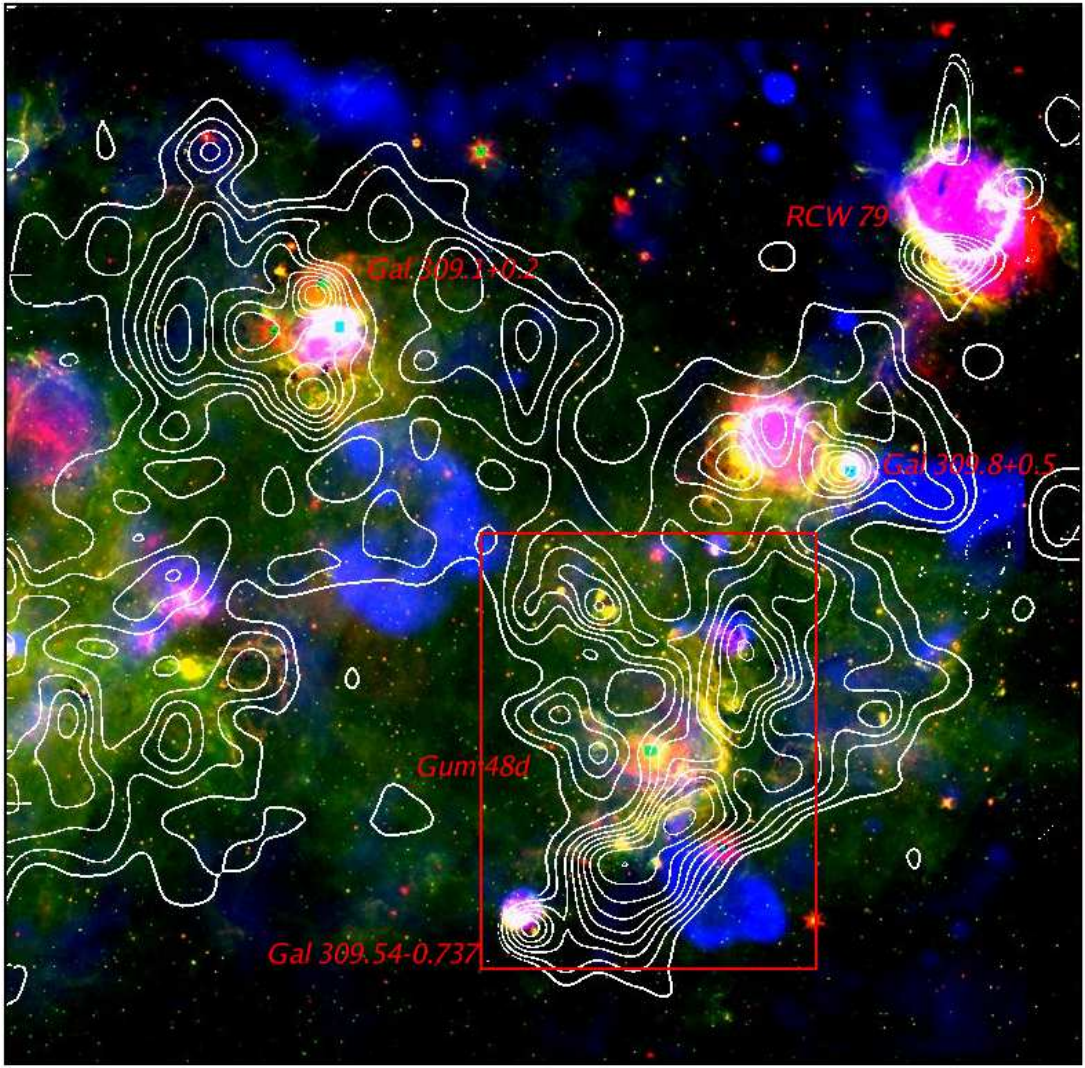}

\caption{Multi-wavelength morphology of the Centaurus Arm.  Blue shows
1420 MHz emission tracing ionized gas (HII regions and SNR), green shows
the 8 micron GLIMPSE image tracing PAH emission (PDR Regions), and red
shows the 24 micron MIPSGAL image, tracing warm dust. The contours show
the location of $^{13}$CO emission at Centaurus Arm velocities, from
\citet{saito01}. The HII regions at Centaurus Arm velocities are
labeled, and the region of Gum48d is marked with a box. 
\label{3col}}
\end{figure}

\begin{figure}
\plotone{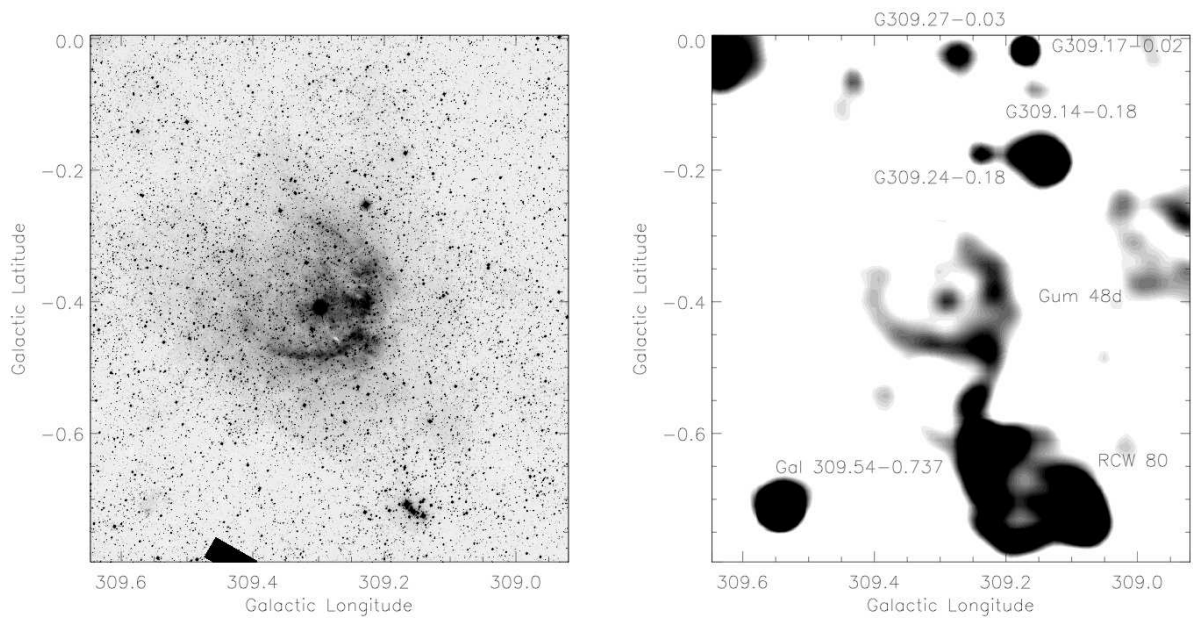}
\caption{Emission from ionized gas in the vicinity of Gum~48d. The
left panel shows the optical emission (H$\alpha$): Gum~48d is clearly
seen as a semi-circular nebulosity. The right panel shows the lower
resolution 1420 MHz radio emission from the same region. The radio
emission is dominated by the SNR RCW~80, but Gum~48d is also clearly
seen. HII regions are labeled. 
\label{ion}}
\end{figure}

\begin{figure}
\plotone{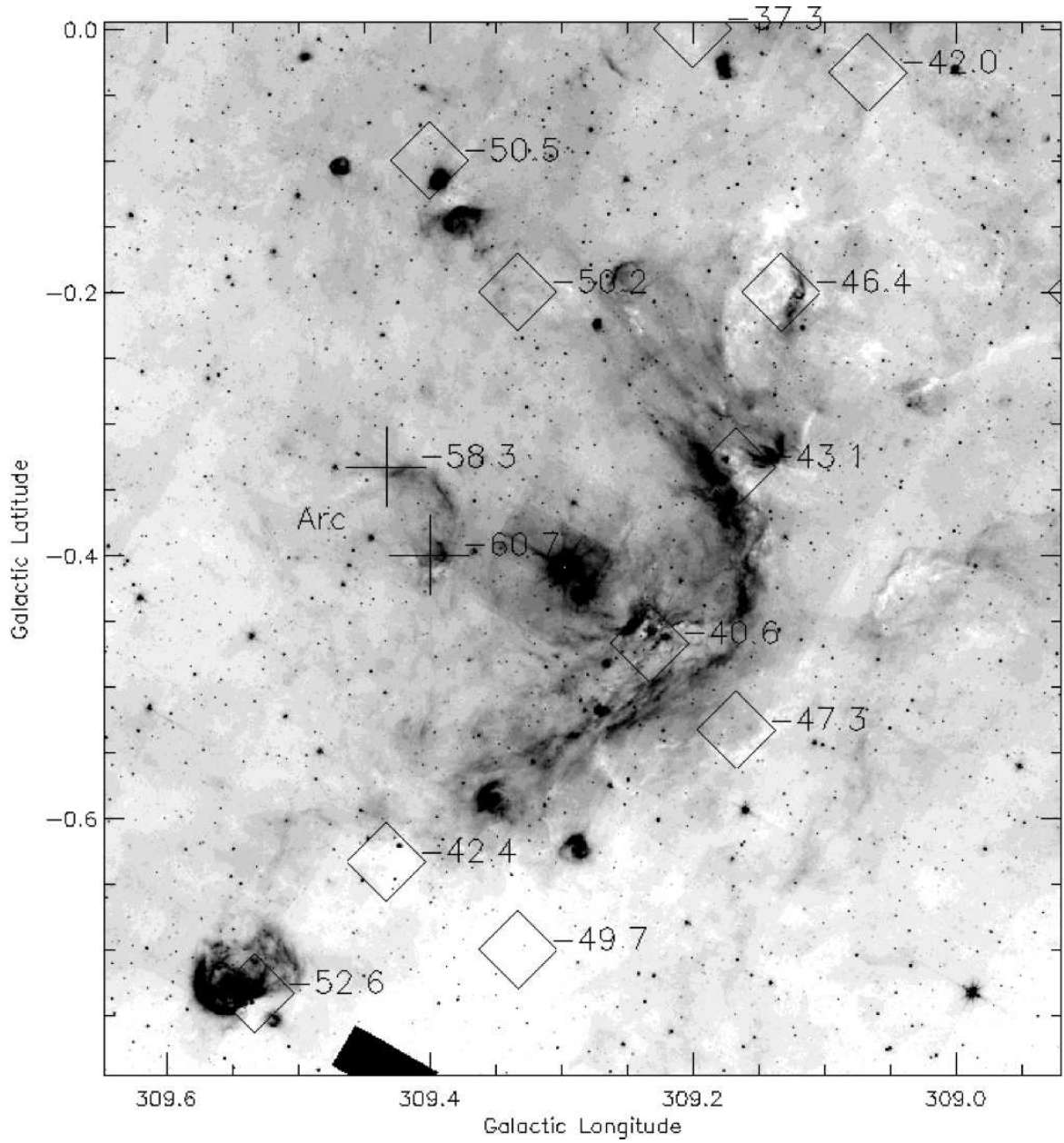}
\caption{Gum 48d at 8 microns (left) tracing PAH emission.  The
positions of molecular clouds at velocities ranging from -36 to -64
km~s$^{-1}$ (the velocity of the Centaurus Arm) are marked, as listed
in \citet{saito01}. Diamonds indicate material in the velocity range
of Gum~48d, crosses those clouds at slightly higher velocities. The
infrared dark clouds can be seen as regions of no emission at 8
microns, here seen as light patches against the diffuse emission.
 \label{pah}}
\end{figure}

\begin{figure}
\plotone{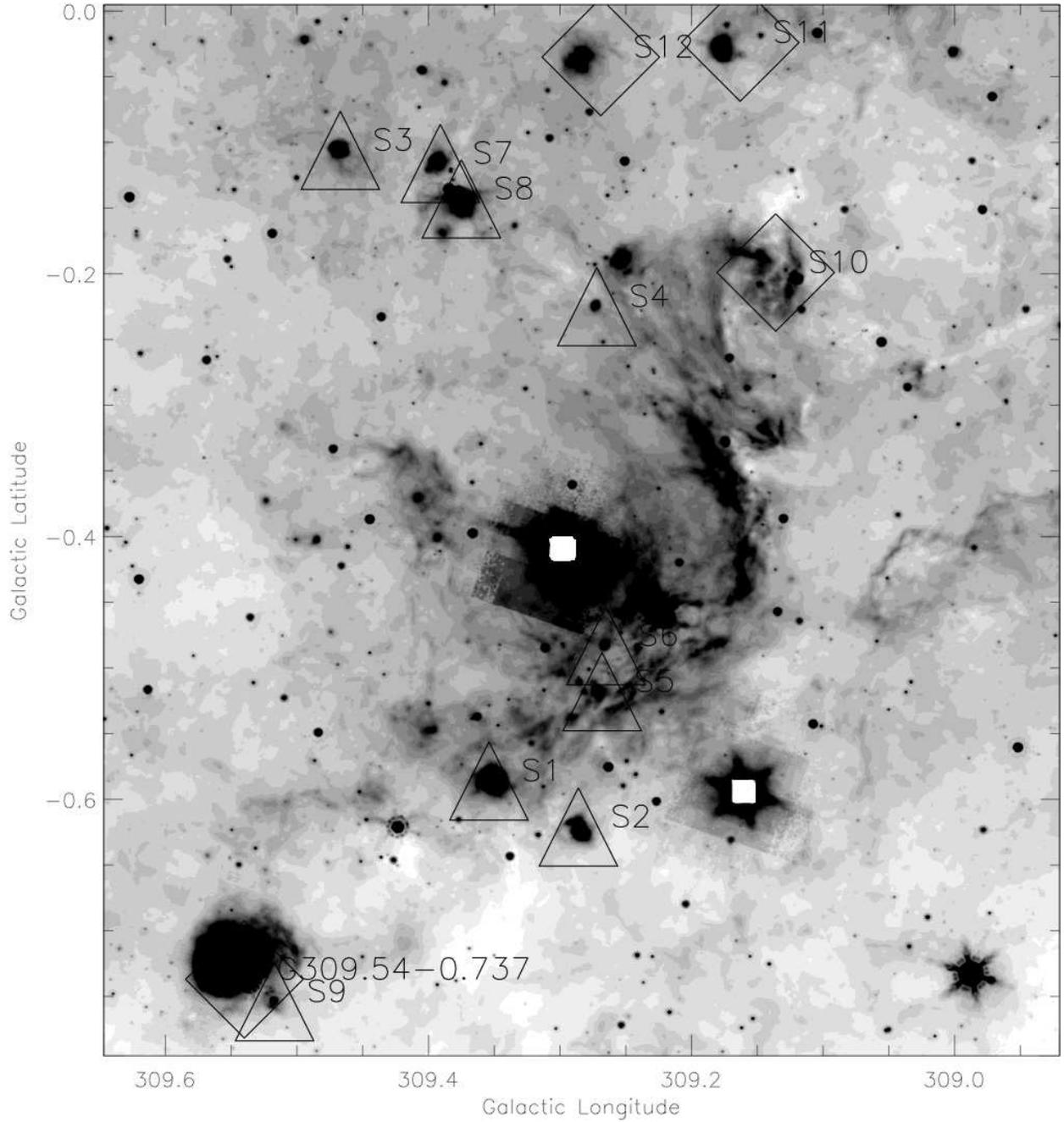}
\caption{Gum 48d at 24 microns (left) tracing emission from warm
dust. Diamonds mark the location of small HII region, while triangles
mark the locations of the potential enshrouded B stars.  The white
squares are saturated pixels.
 \label{warm}}
\end{figure}

\begin{figure}
\plotone{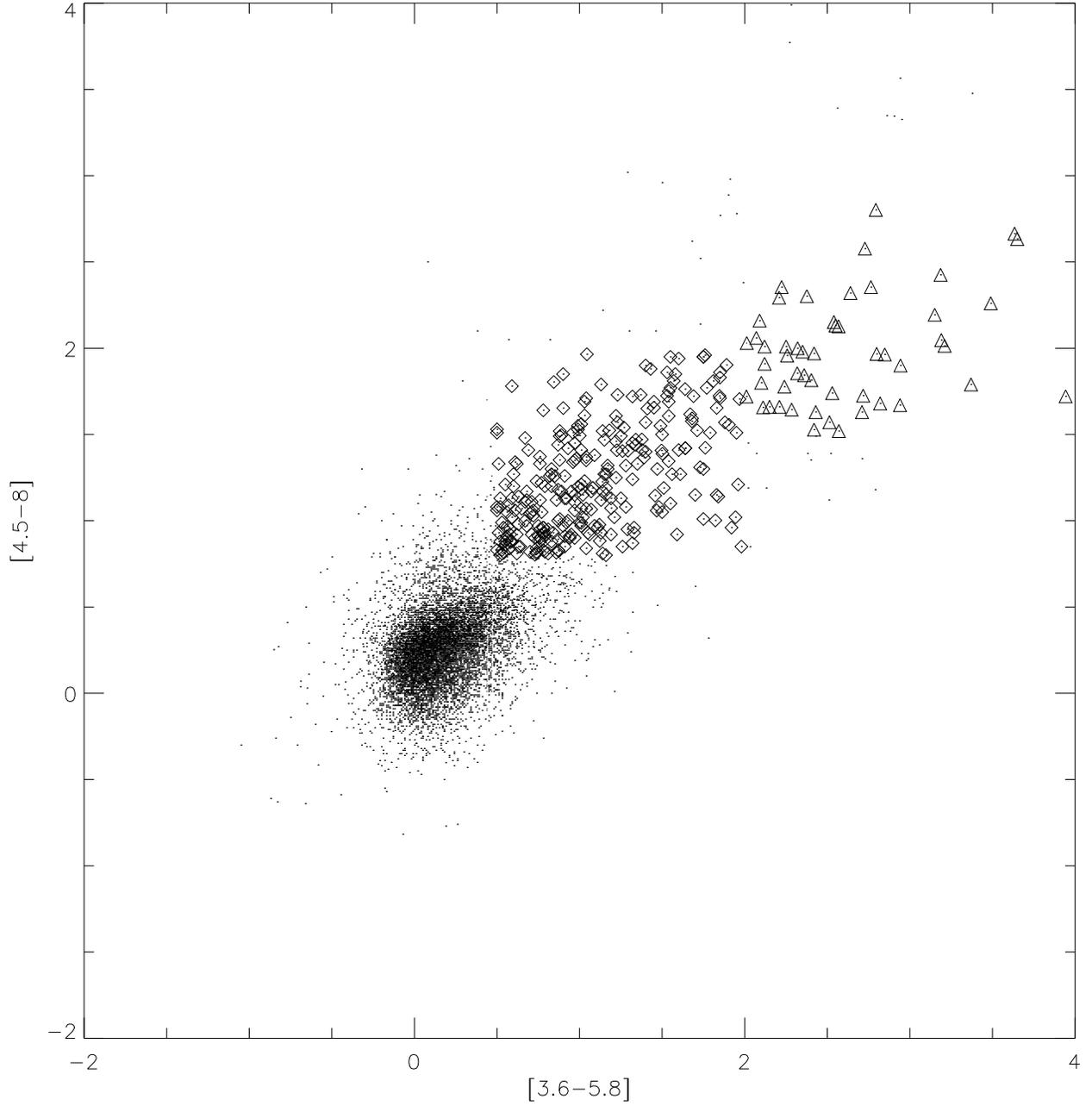}
\caption{IRAC color-color diagram for Gum~48d.  Triangles mark the
Class I sources, diamonds the Class II and points the main sequence
locus.  Flat spectrum sources are not separated out in this technique. 
 \label{cc1}}
\end{figure}

\begin{figure}
\epsscale{0.6}
\plotone{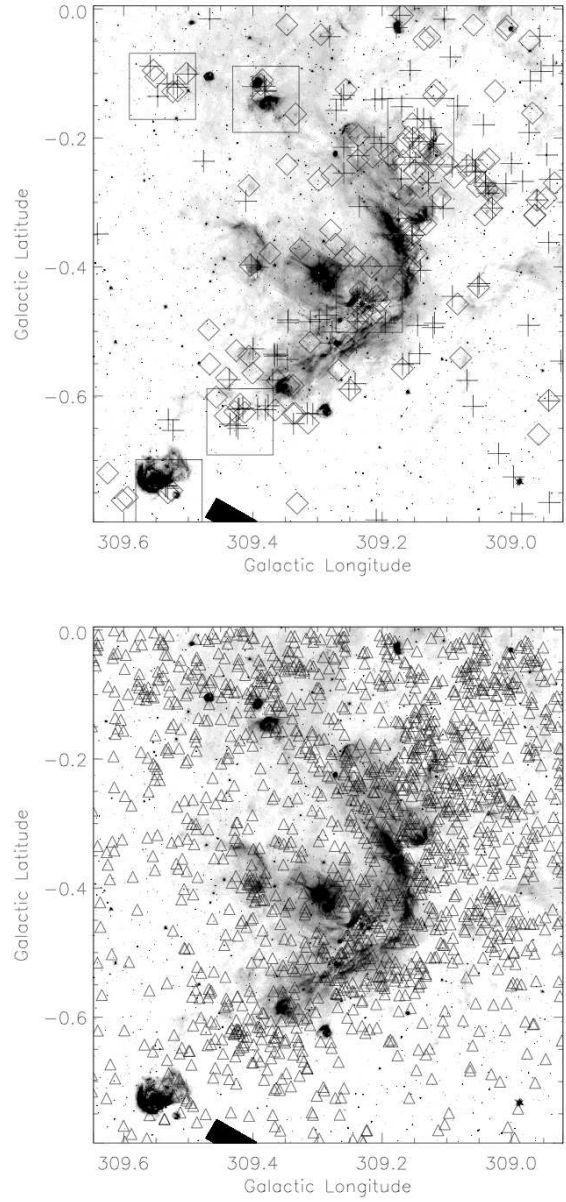}
\caption{Gum 48d shown at 8 microns with star formation
overlayed. Upper panel: Class I sources: crosses, Flat Spectrum:
diamonds, with the location of clusters marked.  Lower panel: class II
sources marked with triangles.
 \label{sform}}
\end{figure}

\begin{figure}
\plotone{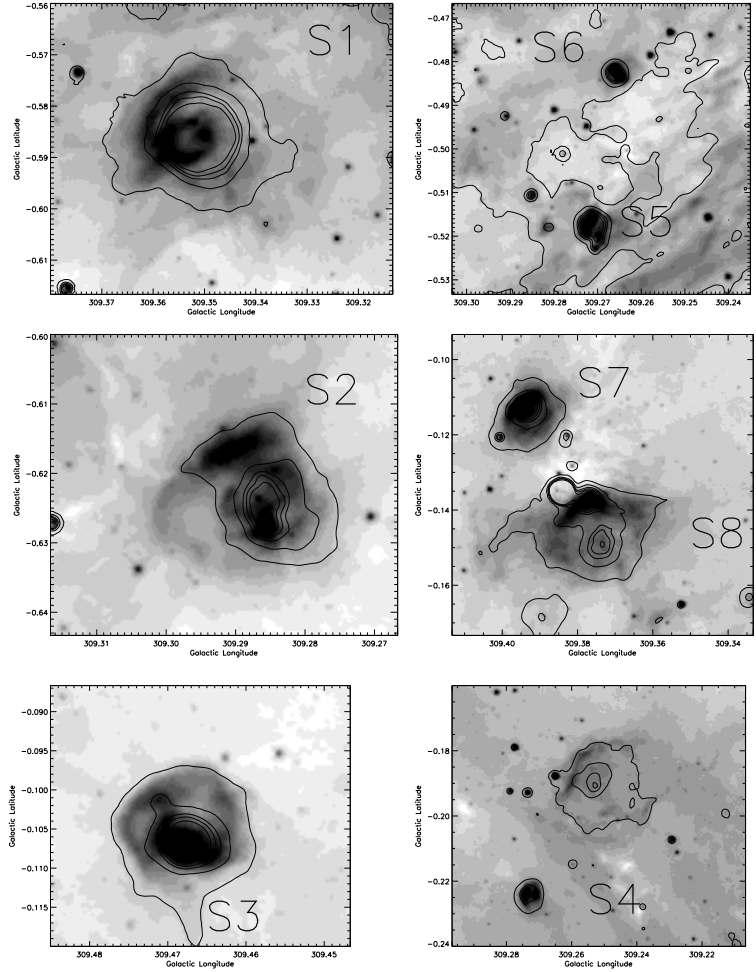}
\caption{A gallery of compact features, shown at 8 microns (grey-scale)
tracing PAH emission and 24 microns (contour) tracing warm dust. The
source numbers are labeled.
 \label{compact1}}
\end{figure}

\begin{figure}
\epsscale{0.50}
\plotone{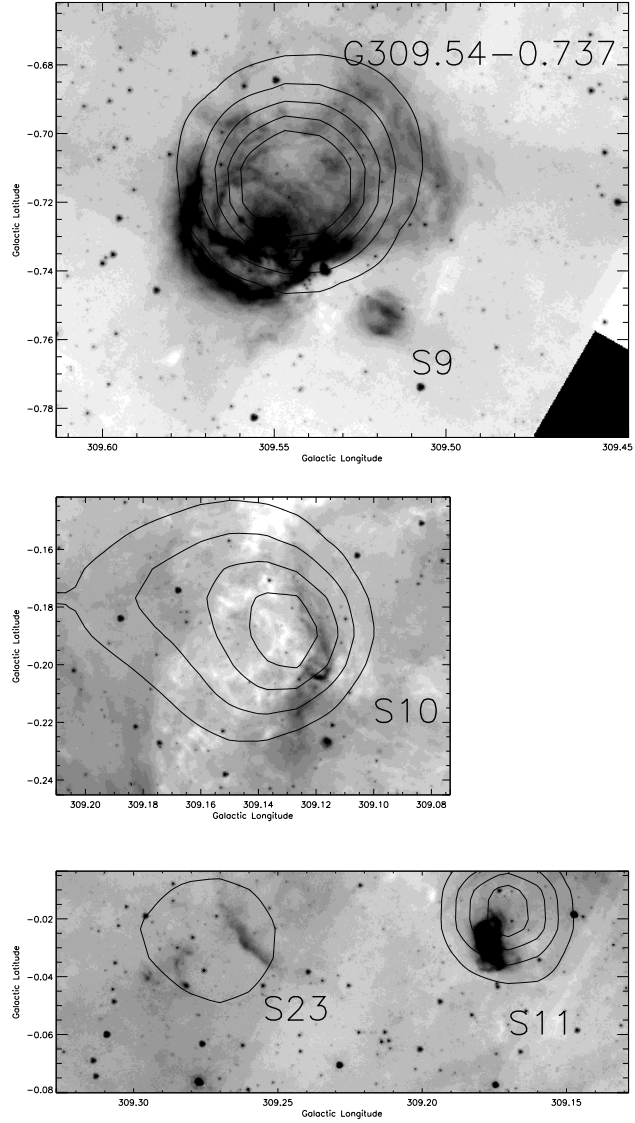}
\caption{Compact features associated with emission from ionized gas.
Greyscale: 8 micron PAH emission, Contour: 1420 MHz emission.  The
source numbers are labeled.
 \label{compact2}}
\end{figure}

\begin{figure}
\epsscale{0.50}
\plotone{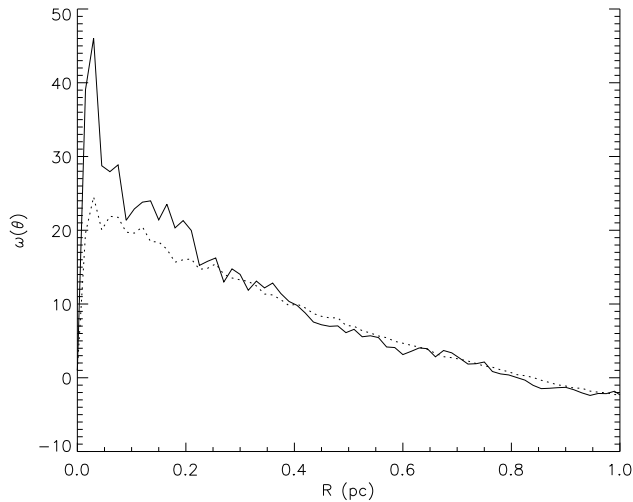}
\caption{A schematic diagram of the history of Gum~48d. (a)
$\approx$10 Myr ago: a giant HII region ionized by multiple OB stars
creates a shell as it expands into giant molecular cloud, (b)
$\approx$4 Myr ago: The giant HII region has consumed much of the
cloud and has accelerated the surrounding molecular material. HR 5171A
and B have formed at the periphery, in the surrounding shell, and have
produced their own HII region, Gum~48d.  (c) $\approx$ 1 Myr ago: The
giant HII region is not longer present, as the ionizing stars have
passed through the SN phase, and what is left is a low density region
with a remnant expanding shell. Gum~48d is in its most energetic
phase. (d) Present: The ionization balance of Gum~48d has changed
considerably, as it is now ionized by a single B0 star. Between the
HII region and the dense molecular cloud is a low density, un-ionized
region, as the recombination timescale is short compared to the age of
the region. This is the remnant of the more energetic HII region.
 \label{sketch}}
\end{figure}

Gum~48d is part of a large molecular complex containing a significant
amount of star formation activity; this region is shown in the three
color image of Figure~\ref{3col}. This molecular complex is one of the
most massive in the galaxy (2.3x10$^{6}$ $\msol$ as traced by
$^{12}$CO(J=1-0)) and forms a massive ridge in CO with velocities
corresponding to the Centaurus Arm \citep{saito01}. The CO ridge is
associated with an even larger ridge in neutral hydrogen.

There are a number of HII regions associated with this ridge.  All of
the HII regions are associated with high mass condensations in the
molecular ridge and have velocities consistent with the Centaurus Arm;
their locations are marked in Figure~\ref{3col}. Gum48d,
Gal~309.1+0.2, Gal~309.8+0.5 and Gal~309.54-0.737 are associated with
the molecular supershell, while RCW~79 is located slightly to the
north-west.  Gal 309.54-0.737 shows the classic morphology of a young
blister HII region, while Gal~309.1+0.2, Gal~309.8+0.5 and RCW~79 show
prototypical main sequence HII region morphologies.

This entire region is studied in more depth in \citet{karr08}. In that
paper, the masses of the ionizing stars needed to maintain the current
1420 MHz luminosity were calculated using the free-free continuum
luminosity.  The \citet{dyson97} model for HII region expansion was
used to estimate the ages of the regions.  The radio luminosities are
consistent with each HII region being ionized by a single OB star
(ranging from O9.5 to O6.5 for all the regions) and the expansion
timescales are consistent with the HII regions being in the 1-3 Myr
age range, with ionizing sources still firmly on the main sequence.

As discussed in Section~\ref{history}, Gum~48d is significantly
different from these regions in both morphology and star formation
content. The emission from ionized gas is faint and diffuse in the
radio, where extinction is not a factor. The morphology of the region
is structured on small scales and the star-formation in the region
includes compact HII regions and enshrouded B-stars in addition to low
mass star formation. The other HII regions regions are bright in the
radio, with clearly defined HII region morphologies and shell like PDR
regions directly bordering on the bright HII regions.  They show
neither the compact features of Gum~48d nor secondary intermediate and
high mass star formation indicative of triggering, although there is
still significant low mass star formation.  On the other hand, they,
contain bright rims, cometary globules and other characteristic
morphological features of the classic HII region which are absent in
Gum~48d.  These differences point to a significantly different 
evolutionary phase of Gum~48d from that of the classic HII regions. 

The proposed history of the star formation over this whole region is
one of multiple, linked generations of star formation, as illustrated
schematically in Figure~\ref{sketch}.  The first generation of stars,
forming 10 Myr ago and consisting of multiple OB stars, culminated in
a series of supernovae explosions 4-6 Myr in the past. The combined
energy of the HII regions, stellar wind and supernovae created an
extensive, massive expanding supershell/ridge around the region, as
discussed in \citet{saito01}.  A second generation of star formation
commenced in the expanding supershell some 1-3 Myr ago, producing the
extended HII regions currently seen in the molecular ridge.  A third
generation of highly embedded star formation is seen around the
individual HII regions, corresponding to star formation triggered by
the HII region itself. This scenario is discussed more fully in
\citet{karr08}.

Gum~48d itself, however, does not fit cleanly into any of the three
phases of star formation. In addition to the morphology and star
formation differences discussed above, it is also of a significantly
different age.  The previous generation and progenitor of the
molecular and neutral supershell/ridge is supposed to have formed
around 10~Myr ago, a factor of 2.5 greater than our derived ages for
the ionizing stars of the region.  For the timescales estimated in
\citet{saito01}, Gum48d could be one of the first of the second
generation of young stars, or possibly one of the last of the
first. The latter option is the more plausible given the relative time
scales.

The motion of Gum~48d's ionizing stars give further support to this
scenario.  The Hipparcos proper motions for HR~5171 are -5.89 mas/year
$\pm$1.38 and -3.66 mas/year $\pm$1.58 for RA and Dec respectively
\citep{perryman97}. This results in a proper motion coming from inside
the large supershell, consistent with the premise that the HR~5171
binary system was formed in the expanding supershell/ridge of material
now seen as the giant molecular cloud.

The molecular material in the vicinity of Gum~48d has a higher column
density than that seen in the vicinity of the other regions (as seen
in Figure~\ref{3col}). This may be an intrinsic property of the
surrounding shell, or may be the result of compression of the existing
material by Gum~48d. If the former, this could explain the earlier
formation of the region, as its more massive stars condensed out of
the densest portions of the molecular ridge.  Being more massive, the
stars progressed through the main sequence more quickly, resulting in
the current striking morphological differences.

Finally, there are some interesting (if purely speculative)
considerations regarding the possibility of triggered star formation
and the associated timescales. Gum~48d is the only region in this
molecular ridge that shows significant evidence for triggered star
formation on intermediate and high mass stars. It is also the oldest
of the regions by 1 to 2 Myr.  Is the timescale for the triggering of
high mass star formation longer than that required for lower mass
stars? Answering this question is beyond the scope of this paper, but
a comparison with other highly evolved, late type HII regions could
prove interesting.

\section{Summary}

\begin{itemize}
\item Gum48d is an old, highly evolved HII region in a post-main
sequence phase that is the remnant of a larger, more energetic main
sequence HII region.  The current morphology of the molecular gas and
PDR region reflects this history, as does the faint remnant HII region
seen in the radio and optical. The region has progressed though the
classic main sequence HII region phase and is now in a late HII region
phase (before the existence of supernovae and a remnant molecular
shell). This represents an interesting and not particularly well
studied phase of HII region evolution.

\item There is significant ongoing star formation associated with
Gum~48d at a variety of masses. Low mass young stars are seen
throughout the molecular cloud, with the more embedded, younger phases
being more highly clustered and associated with the densest molecular
gas.  Intermediate mass star formation, in the form of enshrouded B
stars, and high mass star formation, in the form of compact HII
regions, are seen around the periphery of the molecular/neutral PAH
shell and are similarly clustered.  The timescales of expansion,
fragmentation and star formation are consistent with triggered star
formation by the collect and collapse model.

\item Gum~48d is part of a much larger molecular cloud complex
consisting of a very massive expanding partial supershell/ridge of
molecular and neutral material, the result of a previous generation of
massive star formation 10 Myr ago. The ridge contains a number of HII
regions with ages in the 1-3 Myr range, but Gum~48d is unique in its
older age, diffuse radio morphology, extended PDR and variety of
secondary star formation. While the other regions are clear candidates
for secondary star formation within the expanding molecular
supershell/ridge, Gum~48d is most likely the first and highest mass of
these triggered massive stars, and consequently in a later state
of spectral evolution.

\end{itemize}

This paper makes use of data from the Southern H-Alpha Sky Survey
Atlas (SHASSA), which is supported by the National Science Foundation,
and The Two Micron All Sky Survey, which is a joint project of the
University of Massachusetts and the Infrared Processing and Analysis
Center/California Institute of Technology, funded by the National
Aeronautics and Space Administration and the National Science
Foundation." The Australia Telescope is funded by the Commonwealth of
Australia for operation as a National Facility managed by CSIRO.

\clearpage
\newpage

\clearpage
\newpage

\bigskip

\begin{deluxetable}{lllllllcc}
\tabletypesize{\scriptsize} \tablecaption{Physical parameters of
compact features in Gum~48d. The physical scale assumes a distance of
3.5 kpc.\label{compactparms}} \tablewidth{0pt} \tablehead{

\colhead{Source} & \colhead{glon} & \colhead{glat}&  \colhead{IRAS ID} &
\colhead{Diameter} & \colhead{Diameter} & \colhead{IRAS Flux}
&\colhead{Ionized}\\ \colhead{Number} & \colhead{} & \colhead{} &\colhead{} &
\colhead{arcsec} & \colhead{pc} & \colhead{log($\lsol$))} &
\colhead{Gas}\\} \startdata 

S1  & 309.354 & -0.586 &IRAS 13444-6230 & 69 & 1.3 & 3.52  & \\ 
S2  & 309.286 & -0.621 &IRAS 13439-6233 & 43 & 0.83 & 3.94 & \\ 
S3  & 309.467 & -0.106 &IRAS 13445-6200 & 33 & 0.64 & 3.46 & \\ 
S4  & 309.272 & -0.225 &                & 15 & 0.29 &      & \\ 
S5  & 309.268 & -0.518 &                & 23 & 0.45 &      & \\ 
S6  & 309.265 & -0.483 &                & 32 & 0.62 & 3.77 & \\ 
S7  & 309.391 & -0.116 &IRAS 13439-6207 & 37 & 0.73 & 3.93 & \\ 
S8  & 309.375 & -0.143 &IRAS 13438-6203 & 77 & 1.5 &       & \\ 
S9  & 309.517 & -0.754 &                & 24 & 0.47 &      & \\ 
S10 & 309.140 & -0.194 &IRAS 13418-6210 & 170 & 3.3 &      & G309.14-0.18 \\ 
S11 & 309.176 & -0.030 &IRAS 13419-6159 & 68 & 1.3 &       & G309.17-0.02 \\ 
S12 & 309.273 & -0.035 &                & 99 & 1.9 &       & G309.27-0.03 \\ \enddata
\end{deluxetable}

\end{document}